\documentclass[12pt]{article}
\usepackage{pic02}
\usepackage{hyperref}
\usepackage{url}
\usepackage{graphicx}

\newcommand{\us}{\mbox{$\mu$s}}

\newcommand{\numu}{\mbox{$\nu_{\mu}$}}
\newcommand{\numub}{\mbox{$\bar{\nu}_{\mu}$}}
\newcommand{\nue}{\mbox{$\nu_{e}$}}
\newcommand{\nueb}{\mbox{$\bar{\nu}_{e}$}}
\newcommand{\nutau}{\mbox{$\nu_{\tau}$}}

\newcommand{\ep}{\mbox{e$^{+}$}}

\newcommand{\pos}{\mbox{e$^{+}$}}
\newcommand{\mum}{\mbox{$\mu^{-}$}}
\newcommand{\mup}{\mbox{$\mu^{+}$}}
\newcommand{\pip}{\mbox{$\pi^{+}$}}
\newcommand{\cerenk}{\mbox{\v{C}erenkov}}

\newcommand{\CCprot}{\mbox{p\,(\,\nueb\,,\,\ep\,)\,n }}

\newcommand{\numunue}{\mbox{\numu $\rightarrow\,$\nue }}
\newcommand{\numunutau}{\mbox{\numu $\rightarrow\,$\nutau }}
\newcommand{\numubnueb}{\mbox{\numub $\rightarrow\,$\nueb }}
\newcommand{\NCL}{\mbox{90\%\,C.L.}}

\newcommand{\Dm}{\mbox{$\Delta m^2$}}
\newcommand{\sit}{\mbox{$\sin ^2(2\Theta )$}}

\newcommand{\Gdng}{\mbox{Gd\,(\,n,$\gamma$\,)}}
\newcommand{\pn}{\mbox{p\,(\,n,$\gamma$\,)}}

\begin{document}

\title{\bf SHORT BASELINE ACCELERATOR-BASED NEUTRINO
OSCILLATION SEARCHES}
\author{Klaus Eitel \\
{\em Institut f\"ur Kernphysik, Forschungszentrum Karlsruhe,
76021 Karlsruhe, Germany}\\
{\em e-mail: klaus.eitel@ik.fzk.de}}
\maketitle
\begin{figure}[h]
\begin{center}
%
% include photograph for proceeding version
%
%\includegraphics[height=4.5cm]{eitel.eps}
%
% insert a fixed vertical spacing instead for the ArXiv preprint
%
\vspace{4.5cm}
\end{center}
\end{figure}

\baselineskip=14.5pt
\begin{abstract}
We review the status of the search for neutrino oscillations in
the short baseline regime with experiments at accelerators. The
evidence for \numubnueb\ from the LSND experiment is compared
with the negative results of the KARMEN \numubnueb\ search and
with the results from NOMAD and NuTeV in the same flavor mixing
channel. We describe the upcoming MiniBooNE experiment which
should be sufficiently sensitive to unambiguously confirm or
completely rule out the LSND signal.
\end{abstract}
\newpage

\baselineskip=17pt

\section{Introduction}\label{sec:Introduction}

In the last years, tremendous progress has been achieved to
firmly establish the nature of neutrino oscillations using
neutrinos from the sun \cite{solar} as well as neutrinos produced
in the earth's atmosphere \cite{atmo}. However, with oscillation
parameters accessible to accelerator based experiments, the
situation remains unsettled. There is one evidence for
oscillations in the appearance mode \numubnueb\ from the LSND
experiment which is ruled out in part by other experiments such
as KARMEN, NOMAD and NuTeV.

In this paper, we summarize the status of the search for
appearance of \nueb\ (\nue) from \numub\ (\numu) sources
indicating the flavor oscillation \numubnueb\ (\numunue).
Restricting the discussion to this oscillation channel, we can
simplify the 3-dimensional neutrino mass and mixing scheme to a
2-dimensional one which leads to an oscillation probability $P$
for the appearance of \nueb\ in a pure \numub\ beam of
\begin{equation}
P(\numubnueb) = \sit \cdot \sin^2 \left(\frac{1.27\cdot \Dm \cdot
L}{E}\right) . \label{equ:P-osc}
\end{equation}
As can be seen from equ.~\ref{equ:P-osc}, the experimental
sensitivity to the oscillation parameter \Dm\ is given by the
distance $L$ between source and detector and the neutrino energy
$E$. The ratio $L/E$ in the above formula is typically given in
units of m/MeV (see section \ref{sec:MEsources} for LSND and
KARMEN or km/GeV (see section \ref{sec:HEsources} for NOMAD and
NuTeV).

\section{Medium-energy beam stop sources}\label{sec:MEsources}

Existing medium-energy high intensity neutrino sources
(LANSCE\footnote{Los Alamos Neutron Science Center, Los Alamos,
USA}, ISIS\footnote{ISIS at Rutherford Appleton Laboratory,
Chilton, UK}) are based on accelerators with a proton kinetic
energy of 800\,MeV and a beam stop or neutrino production target
of high-Z material. Due to the massive target (e.g. Copper,
Tantalum or Uranium), the resulting neutrino fluxes arise from
\pip\ and \mup\ decays at rest (DAR) and hence are well
understood. $\pi^-$ and $\mu^-$ that stop are readily captured in
the high-Z material of the beam stop. The production of kaons or
heavier mesons is negligible at these proton energies. The \nueb\
flux is calculated to be only $< 8 \cdot 10^{-4}$ as large as the
\numub\ flux, so that the observation of a \nueb\ event rate
significantly above the calculated background would be evidence
for \numubnueb\ oscillations.

\begin{figure}[htb] \begin{center}
\includegraphics[width=12cm]{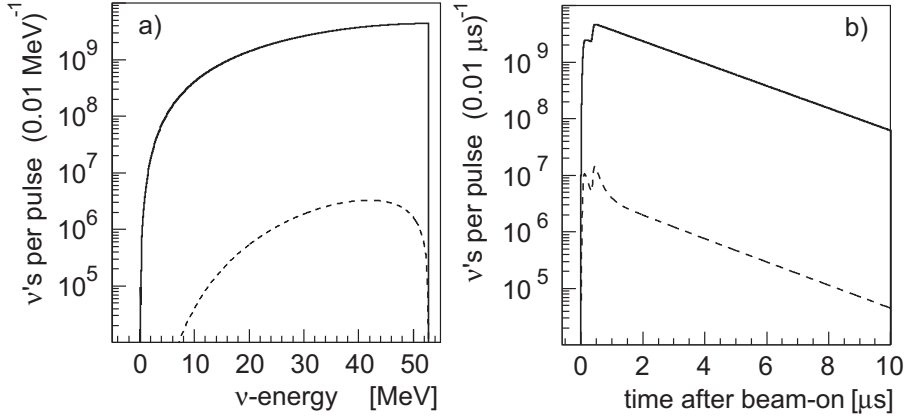}
 \caption{\it (a) Energy spectrum of \numub\ from \mup\ DAR (solid line)
  and of the \nueb\ intrinsic contamination from \mum\ DAR (dashed) for a
  massive Tantalum beam stop target.
  (b) Time distribution of neutrinos after protons on target for
  the ISIS target.
 \label{fig:KARMENflux} }
\end{center} \end{figure}
Figure~\ref{fig:KARMENflux}(a) shows the \numub\ spectrum from
\mup\ DAR (solid line) together with the \nueb\ contribution. The
absolute flux corresponds to the neutrino production of the ISIS
accelerator with 200\,$\mu$A proton current and a Ta--D$_2$O
target. If the time structure of the proton pulses is
significantly smaller than the muon lifetime of 2.2\,\us, the
neutrinos show the corresponding exponential decrease (as in
Fig.~\ref{fig:KARMENflux}(b), an important feature to suppress
cosmic induced background). Otherwise, as for LANSCE with its
600\,\us\ long proton pulses, neutrinos are almost constantly
produced during beam-on times.

\subsection{The LSND evidence for \numubnueb }\label{subsec:LSND}

The LSND experiment took data over the years 1993--1998. During
this period the LANSCE accelerator delivered 28\,896\,C of
protons on the production target. The LSND detector consisted of
an approximately cylindrical tank of $\approx$210\,m$^3$ volume.
The center of the detector was 30\,m from the beam stop neutrino
source. The tank was filled with liquid scintillator of low
scintillator concentration which allowed the detection of both
\cerenk\ and scintillation light. For all details see
\cite{LSNDfinal} and references therein.

A \nueb\ signal from \numubnueb\ oscillations consists of a
spatially correlated delayed (\pos,n) sequence from \CCprot . This
requires a prompt 'electron-like' event with energy $E_e>20$\,MeV
followed by a low energy $\gamma$--event from neutron capture. The
information about the delayed event is compressed into a
likelihood ratio $R_\gamma$: If within one millisecond after the
initial event another event is recorded at distance $\Delta r\le
250$\,cm, the ratio of likelihood $R_\gamma$ in energy, time and
distance of being a correlated \pn\ over an accidental
coincidence is calculated, otherwise $R_\gamma=0$.
\begin{figure}[htb] \begin{center}
\includegraphics[width=11cm]{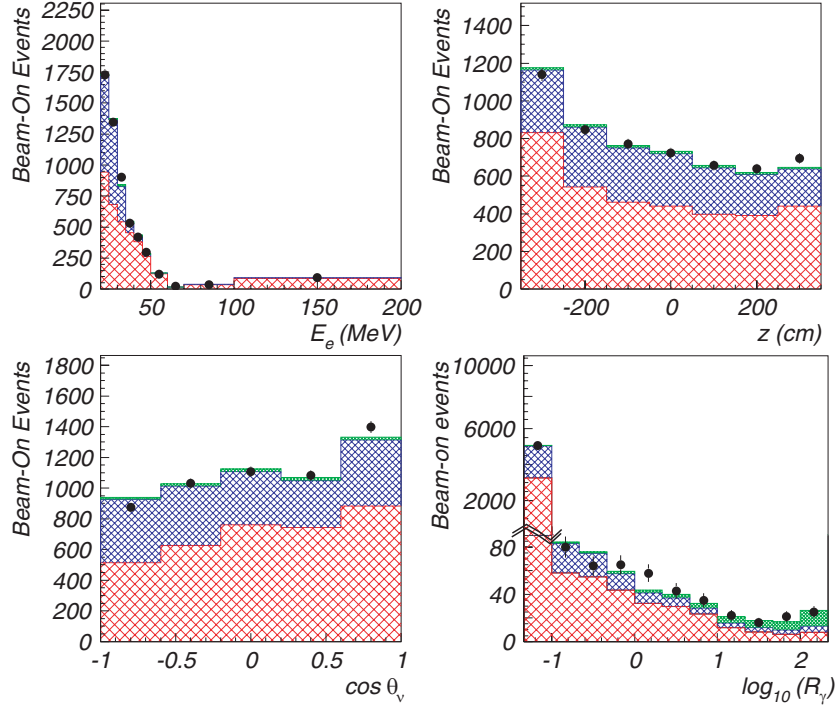}
 \caption{\it LSND data sample with 5697 beam-on events after applying
  loose cuts used for the maximum likelihood analysis (details see text).
 \label{fig:LSNDdata} }
\end{center} \end{figure}
To determine the oscillation parameters \sit\ and \Dm\ in a
likelihood analysis, only loose cuts are applied. The beam-on
event sample then contains 5697 candidates shown in
Fig.~\ref{fig:LSNDdata} with their spectral information in visible
energy $E_e$, spatial distribution along detector axis $z$, the
angle between the incident neutrino direction and the visible
electron-like track $\cos\theta_{\nu}$ and the (\pos,n)
correlation parameter $R_\gamma$.

The event-based lhd function reaches its maximum at $(\sit,
\Dm)_{max} = (0.003, 1.2 \rm{eV}^2)$ corresponding to an excess
of 89.5 events attributed to oscillations \numubnueb . With a
difference of $lnL_{max}-lnL_{no osc.}=14.5$, the oscillation
hypothesis is strongly favored over the no--oscillation case. The
contributions in Fig.~\ref{fig:LSNDdata} (from bottom to top)
show the result of a likelihood analysis in all 4 parameters:
3664.6 cosmic induced events, 1936.8 events induced by known SM
processes and on top (light grey) the about 90 excess events
\footnote{The sum of these 3 contributions leaves 6.1 events
attributed to the appearance of \nue\ from \numunue\ with \numu\
arising from \pip\ decays in flight. See \cite{LSNDfinal} for
details.} from \numubnueb . We will discuss the results in terms
of the oscillation parameters later in section \ref{subsec:joint}.

\subsection{The KARMEN limit for \numubnueb }\label{subsec:KARMEN}

The KARMEN experiment used the ISIS neutrino source. In contrast
to the LANSCE source, the protons are extracted from the
synchrotron as an intense but narrow double pulse of 525\,ns
total width. The unique time structure of the ISIS proton pulses
allowed a clear separation of $\nu$--induced events from any beam
unrelated background. The KARMEN detector was a rectangular high
resolution liquid scintillation calorimeter, located at a mean
distance of 17.7\,m from the ISIS target and shielded by a
multilayer active veto system and 7000\,t steel. The liquid
scintillator volume (65\,m$^3$) was optically separated into 512
independent central modules. Gadolinium was implemented between
the module walls for an efficient detection of thermal neutrons
\Gdng\ with on average 3 $\gamma$'s of energy $\sum
E_\gamma=8$\,MeV.

The KARMEN\,2 experiment took data from February 1997 to March
2001. During this time, protons equivalent to a total charge of
9425 Coulombs have been accumulated on the ISIS target.

\begin{figure}[htb] \begin{center}
\includegraphics[width=11cm]{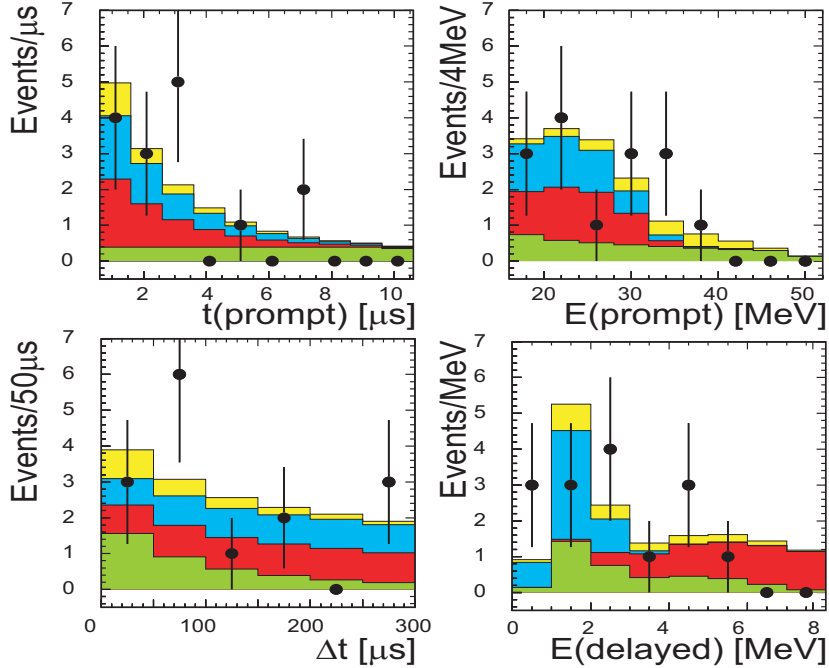}
 \caption{\it KARMEN 15 coincidence candidates after final cuts.
 \label{fig:KARMENdata} }
\end{center} \end{figure}

\begin{figure}[htb] \begin{center}
\includegraphics[width=12cm]{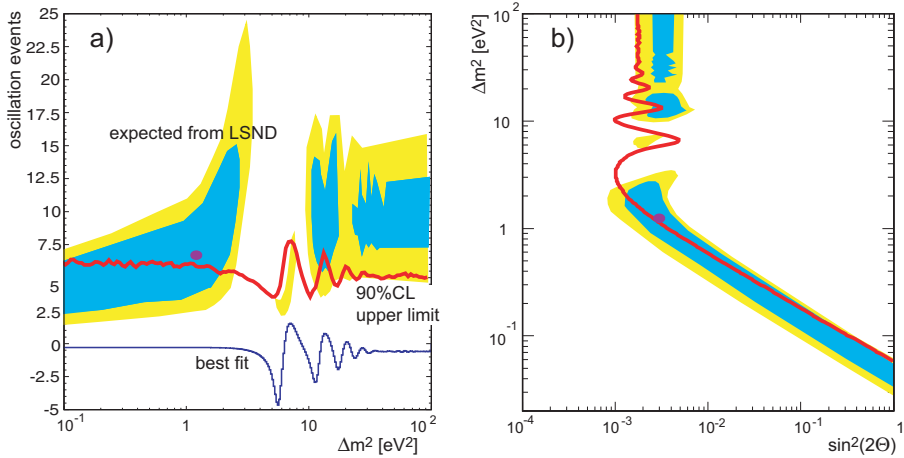}
 \caption{\it (a) KARMEN limit and expected event rate following the
 LSND signal strength. (b) oscillation plot with LSND favored
 regions and KARMEN exclusion curve. The dot denotes the LSND best
 fit values.
 \label{fig:LK2res} }
\end{center} \end{figure}

As for LSND, a \nueb\ signal from \numubnueb\ oscillations
consists of a spatially correlated delayed (\pos,n) sequence. The
requirements for event sequences in KARMEN are described in
detail in \cite{K2paper} and references therein. Applying all
cuts to the data, 15 (\pos ,n) candidate sequences were finally
reduced. Figure~\ref{fig:KARMENdata} shows the remaining sequences
in the appropriate energy and time windows.
The background components are also given with their
distributions. All components except the intrinsic \nueb\
contamination are measured online in different time and energy
windows (see Table~\ref{tab:K2bgsum}).
\begin{table*}
\centering
  \caption{\it Expected KARMEN\,2 background components and \numubnueb\ signal for
  $(\sit=1,\Dm=100\,{\rm eV}^2)$}
  \label{tab:K2bgsum}
  \begin{tabular}{lcl}
   \hline Process & Expectation & Determination\\  \hline
   Cosmic induced background & $3.9 \pm 0.2$ &  meas. in diff. time window\\
   Charged current coinc. & $5.1 \pm 0.2$ &  meas. in diff. energy, time wdws.\\
   \nue --induced random coinc. & $4.8 \pm 0.3$ &  meas. in diff. time wdw.\\
   \nueb\ source contamination & $2.0 \pm 0.2$ &  MC--simulation\\ \hline
   Total background $N_{bg}$ & $15.8 \pm 0.5$&   \\ \hline
   $N_{osc}$ & $2913\pm 267$ & \\ \hline
  \end{tabular}
\end{table*}
The extracted number of sequences is in excellent agreement with
the background expectation, consistent with no oscillation
signal. To also include the detailed spectral information of each
individual event, an event-based maximum likelihood method is
applied. Fig.~\ref{fig:LK2res}(a) shows the results of this
analysis in terms of oscillation events as function of \Dm . The
best fit line is compatible with zero. Applying a unified
frequentist approach \cite{feldcous} leads to the given \NCL\
upper limit which can be compared to what one would expect as
signal strength in KARMEN taken the evidence from LSND.

\subsection{LSND and KARMEN: a critical review}\label{subsec:joint}

Fig.~\ref{fig:LK2res} shows a comparison of the LSND and KARMEN
results obtained after analysing the final data samples with an
event-based maximum likelihood method. In
plot~\ref{fig:LK2res}(a), the LSND evidence with its statistical
error is translated into an expected range of oscillation events
among the KARMEN data. At large values of \Dm , the experimental
outcomes are in clear contrast whereas at small \Dm , the
expected event number is below the \NCL\ exclusion curve.
Fig.~\ref{fig:LK2res}(b) shows the same results, the KARMEN
exclusion curve (for large \Dm\  $\sit < 1.7 \cdot 10^{-3}$ \NCL)
as well as the evidence band of LSND ($lnL_{max}-2.3$,
$lnL_{max}-4.6$), now as functions of the oscillation parameters
\sit\ and \Dm .

Since both experiments have similar sensitivity to the
oscillation parameters but different central statements about
oscillations leading to a partial overlap of exclusion curve and
favored region, one has to solve the following problems
quantitatively: What is the level of compatibility of both
experiments? Assuming statistical compatibility, what are the
oscillation parameters accepted by both experiments?
\begin{figure}[htb] \begin{center}
\includegraphics[width=12cm]{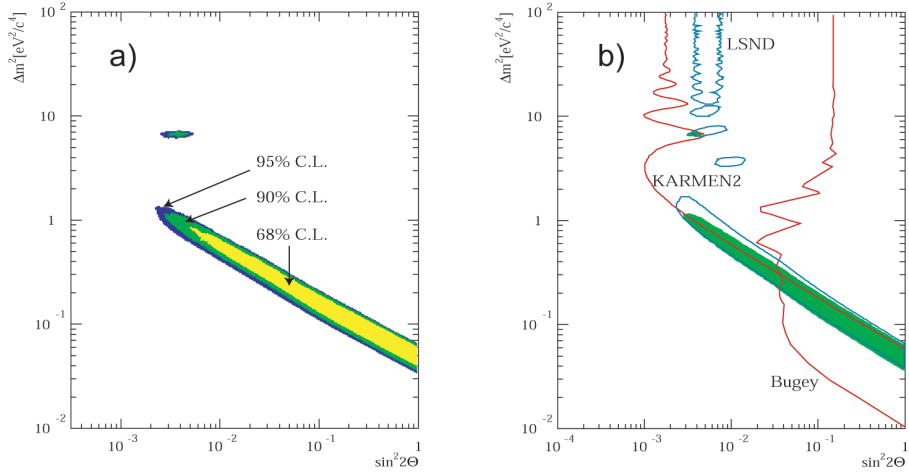}
 \caption{\it (a) Common parameters at various levels of confidence in case of
 KARMEN-LSND compatibility. (b) Region of 58\,\%
 total confidence compared with results of the individual experiments.
 \label{fig:LK2joint} }
\end{center} \end{figure}
A detailed combined statistical analysis \cite{joint} has been
performed to answer these questions. We summarize the main results
also shown in Fig.~\ref{fig:LK2joint}. At a combined level of
60\,\%\,$\times$\,60\,\%\,=36\,\%, the LSND and KARMEN results are
incompatible. Taking the rest probability of 64\,\%, the favored
parameter regions are shown in plot~\ref{fig:LK2joint}(a). In
Fig.~\ref{fig:LK2joint}(b), the parameter region of
90\,\%\,$\times$\,64\,\%\,=58\,\% probability of occurrence is
shown together with the individual results obtained in the same
analysis \cite{joint}.

To summarize, LSND and KARMEN are incompatible at a level of
36\,\% confidence. Assuming statistical compatibility, all
parameter combinations with $\Dm > 1$\,eV$^2$ are excluded apart
from a little 'island' at $\approx 7$\,eV$^2$.

\section{Experiments with neutrino beams}\label{sec:HEsources}

In section~\ref{sec:MEsources}, we dealt with neutrino energies
of tens of MeV. Neutrinos from high energy proton beams typically
have energies of tens up to hundreds of GeV. These higher
energies increase the cross section for neutrino interactions
with the detector material as well as decrease substantially any
beam-uncorrelated background. However, the neutrino flux has to
be modeled carefully and cannot --as for the beam stop neutrino
source-- simply calculated analytically. Detectors are typically
at a distance of several hundreds of meters from the neutrino
source, leading to a so-called short baseline regime: $L/E\approx
0.5/20$\,km/GeV corresponding to a very high sensitivity in \Dm\
for values larger than several tens of eV$^2$ mainly limited by
the intrinsic 'wrong' flavor component within the beam, typically
of the order of $\nueb/\numub(\nue/\numu)\approx$\,1--2\,\%.

\subsection{The NOMAD \numunue\ search}\label{subsec:NOMAD}

The NOMAD experiment took data from 1995--1998 using the CERN wide
band neutrino beam from the 450\,GeV PS. We summarize here only
the search for \numunue\ and refer the reader for all details of
the experiment as well as the \numunutau\ search to
reference~\cite{NOMADmutau}. The MOMAD detector was a high
resolution detector to separate $\nu$ induced charged current
(CC) events by kinematic criteria with a momentum resolution of
3.5\,\% ($p<10$\,GeV). It had a fine grained calorimetry leading
to $\Delta E/E = 3.2\,\%/\sqrt{E}\oplus 1$\,\% and a particle
identification via a TRD resulting in a pion rejection of $10^3$
with electron efficiency $>90$\,\%.
\begin{figure}[htb] \begin{center}
 \includegraphics[width=12cm]{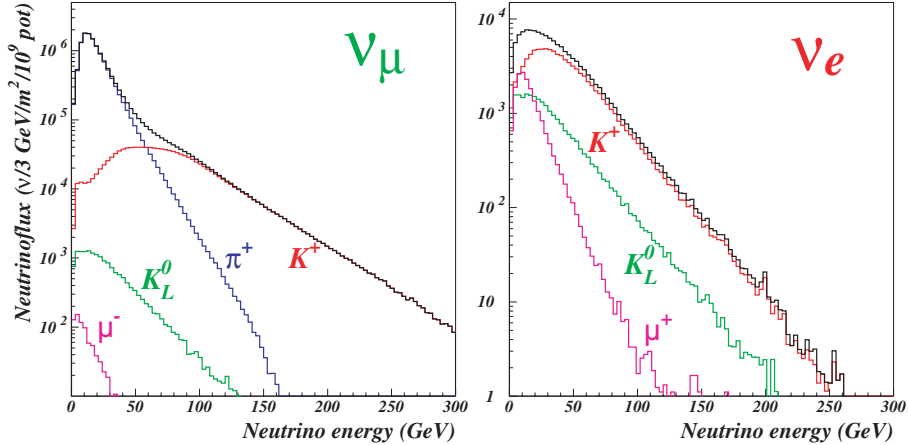}
 \caption{\it NOMAD \numu\ and \nue\ fluxes from various decay modes.
 \label{fig:nomadfluxes} }
\end{center} \end{figure}
Fig.~\ref{fig:nomadfluxes} shows a simulation of the \numu\ and
\nue\ spectra within a transverse fiducial area of $260\times
260$\,cm$^2$ of the detector \cite{NOMADmue}.
\begin{figure}[htb]
 \begin{center}
 \includegraphics[width=14cm]{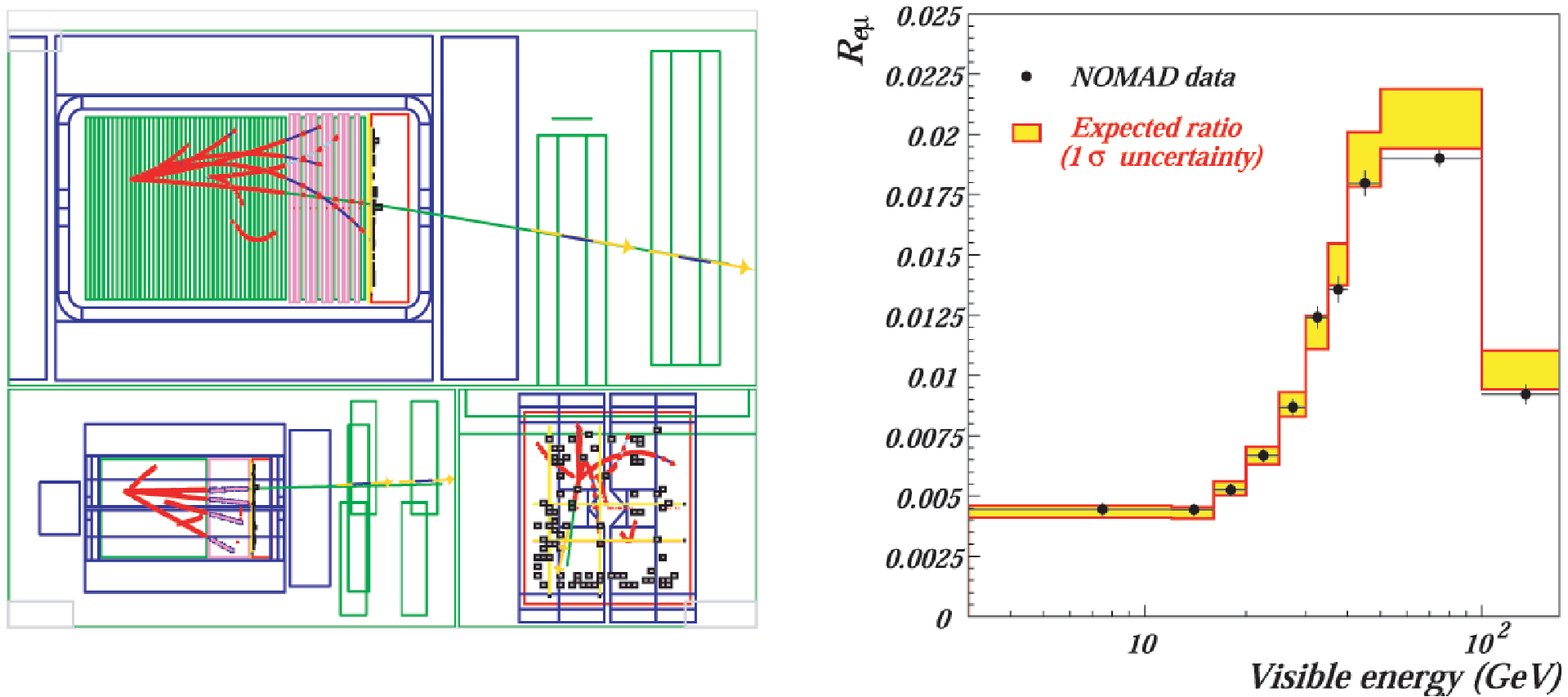}
 \caption{\it A typical \numu --CC event in NOMAD (left);
 ratio $R_{e\mu}$ of \nue/\numu\ --induced CC events.
 \label{fig:nomadresults} }
\end{center} \end{figure}
The NOMAD \numunue\ search was an appearance experiment based on
electron identification with \nue\ contamination in the \numu\
beam of $\nue/\numu \approx 1.5$\,\%. The different energy spectra
and radial distributions of \nue\ and \numu\ within the detector
were used to calculate an expected ratio $R_{e\mu} = (\nue\ {\rm
CC})/(\numu\ \rm{ CC})$ of $\nu$--induced events as function of
$E_{\nu}$ and $r$ (see Fig.~\ref{fig:nomadresults}(left) for a
typical \numu--induced CC event). This ratio was then compared
with the reconstructed events containing 748000 \numu\ CC and 8000
\nue\ CC candidates. The $R_{e\mu}$ distribution in data and MC
(Fig.~\ref{fig:nomadresults}) shows no excess of \nue\ CC,
therefore leading to an exclusion of oscillation parameters, i.e.
for large \Dm\ $\sit < 1.2 \cdot 10^{-3}$ at \NCL\
\cite{NOMADmue} (see also Fig.~\ref{fig:osciall} for the
preliminary NOMAD \numunue\ exclusion curve). It is interesting
to note that the NOMAD \numunue\ result underlines the KARMEN
exclusion at values $\Dm>10$\,eV$^2$ and that the \NCL\ exclusion
curve narrowly passes the allowed 'island' at $\Dm \approx
7$\,eV$^2$.

\subsection{The NuTeV \numunue\ and \numubnueb\ searches}\label{subsec:NuTeV}

The NuTeV experiment at Fermilab took data from 1996--1997, using
the 800\,GeV primary proton beam from the Tevatron. A sign
selected quadrupole train lead to a very effective separation of
neutrino and anti-neutrino beams, allowing separate searches of
\numunue\ and \numubnueb\ oscillations by appearance of \nue\ or
\nueb, respectively. After a 320\,m decay pipe followed by 915\,m
of shielding, the detector consisted mainly of a sandwich system
of steel slabs and scintillator planes. CC events induced by
\numu\ or \nue\ can be separated via the energy deposit in the
first 3 scintillators compared to the total deposit $\eta 3 =
(E_1+E_2+E_3)/E_{\rm cal}$: 'short' tracks indicate electron like
events whereas 'long' tracks point to muons. For all details of
the oscillation search see \cite{NuTeV} and references therein
for the experimental setup.
\begin{figure}[htb] \begin{center}
\includegraphics[width=14cm]{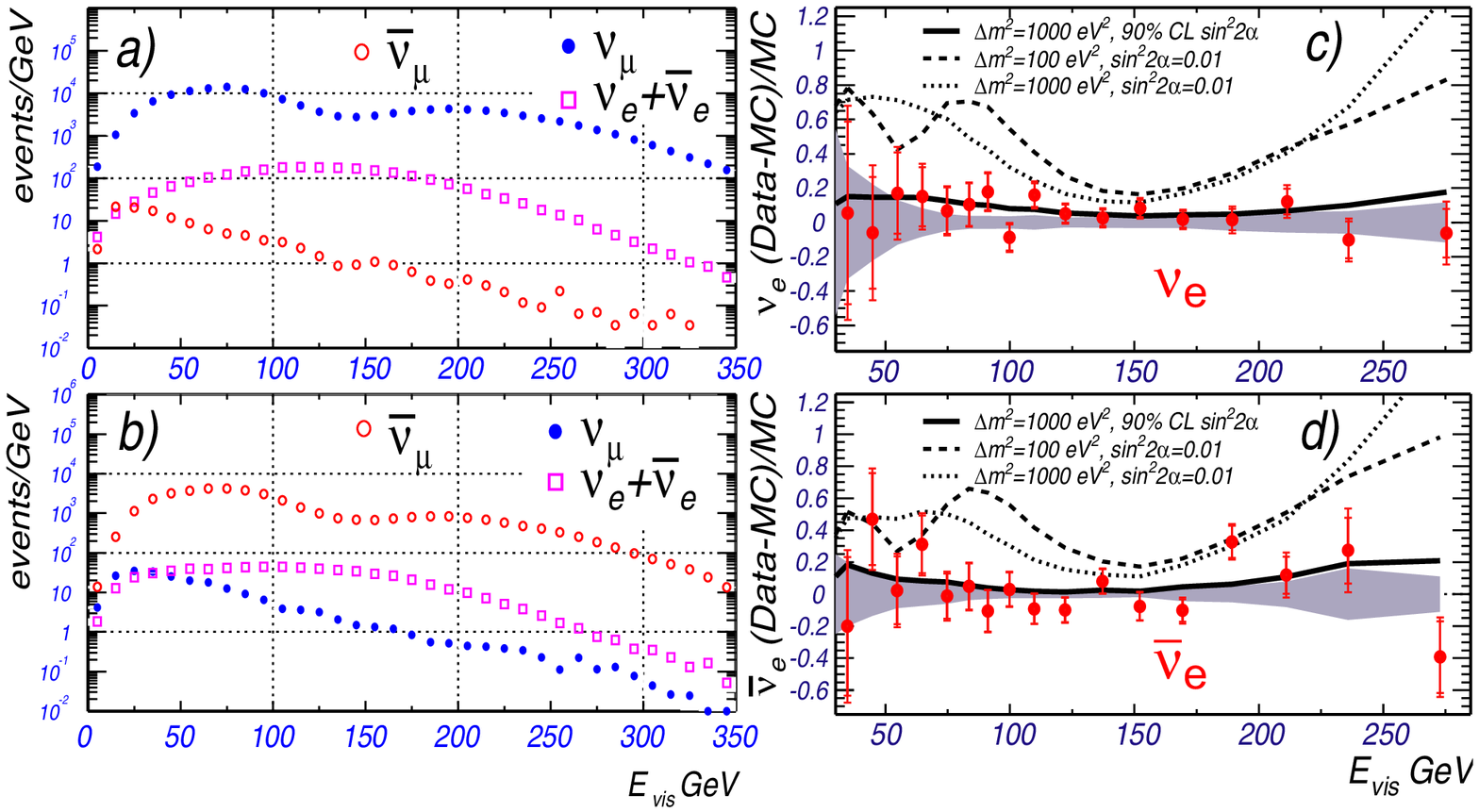}
 \caption{\it Visible energies of neutrino (a) and anti-neutrino (b) fluxes
              at NuTeV; ratio of measured to expected \nue\ (c) and \nueb\
              (d) induced events.
 \label{fig:nutevfluxes} }
\end{center} \end{figure}

A detailed beam simulation predicts the visible energy spectra
from various $\nu$--induced CC events given in
Fig.~\ref{fig:nutevfluxes}(a) and (b).
Fig.~\ref{fig:nutevfluxes}(c) and (d) show the ratios of the
detected over predicted (\nue, \nueb) event numbers versus visible
energy minus 1. The lines correspond to the predictions of various
oscillation scenarios, the solid is the \NCL\ upper limit for
$\Dm=1000$\,eV$^2$. The exclusion curve for \numubnueb\ as
function of the oscillation parameters relevant for LSND is shown
in Fig.~\ref{fig:osciall} excluding the very high \Dm\ region
thereby confirming KARMEN and NOMAD.

\subsection{The MiniBooNE experiment}\label{subsec:BooNE}

The MiniBooNE experiment at Fermilab is designed to cover the
entire parameter range favored by LSND with a completely
different experimental setup \cite{BooNE} by looking for \numu\
disappearance and \nue\ appearance from \numunue\ in a \numu\
beam.
\begin{figure}[htbp]
  \centerline{\hbox{
    \includegraphics[width=6.5cm]{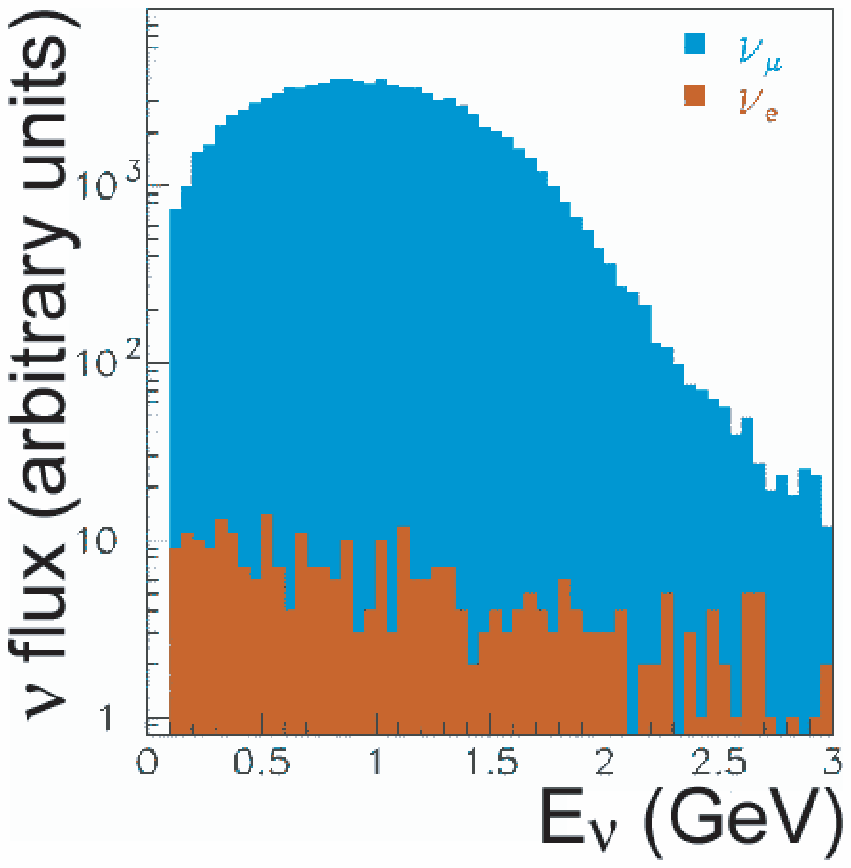}
    \hspace{0.3cm}
    \includegraphics[width=7cm]{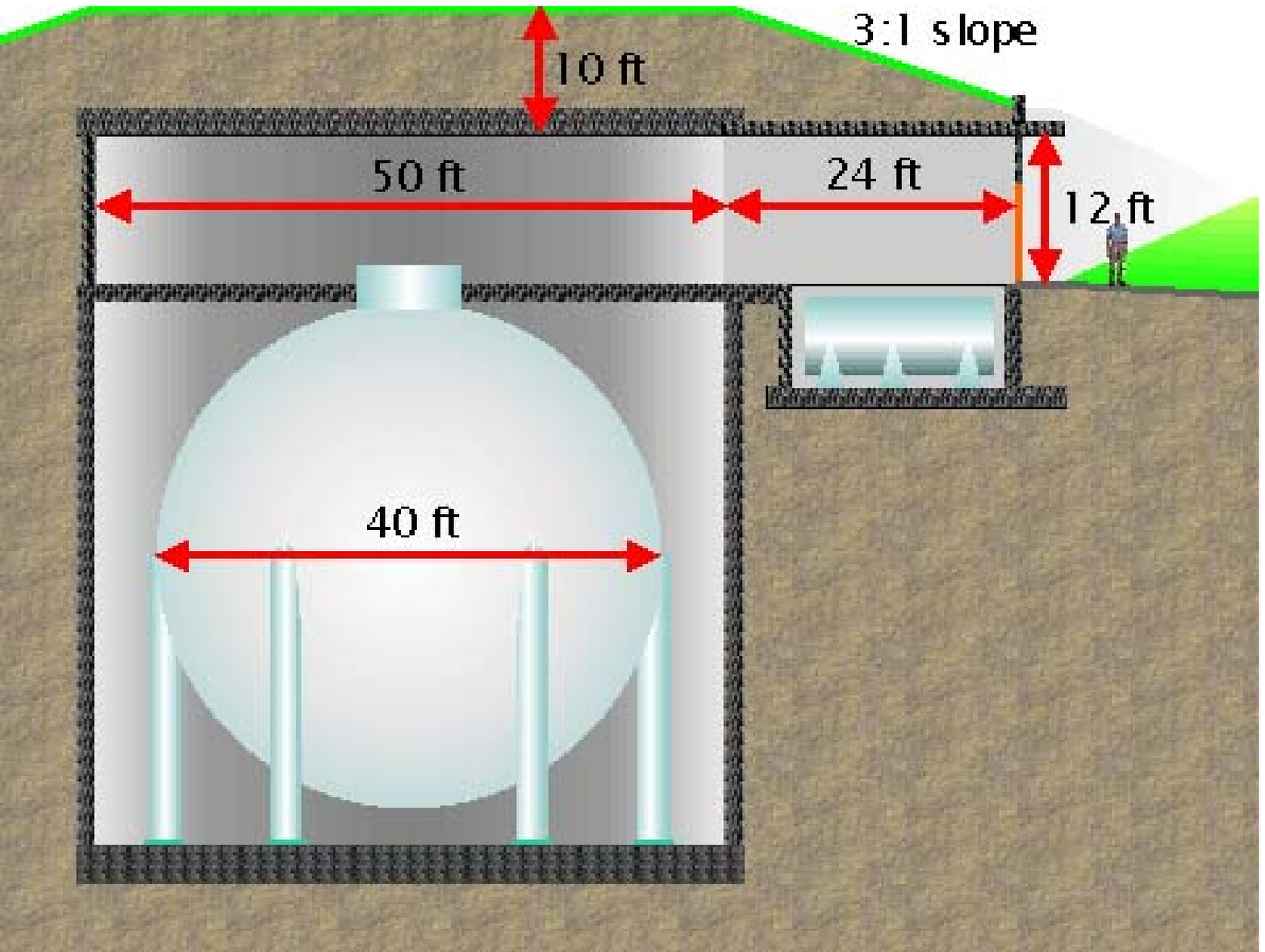}
    }
  }
 \caption{\it MiniBooNE expected neutrino flux and schematic view of the detector site.
 \label{fig:booneflux} }
\end{figure}
%\begin{figure}[htb] \begin{center}
%\includegraphics[width=14cm]{MiniBooNE-flux.eps}
% \caption{\it MiniBooNE expected neutrino flux and schematic view of the detector site.
% \label{fig:booneflux} }
%\end{center} \end{figure}
MiniBooNE will use the 8\,GeV booster at Fermilab's main injector
to produce a neutrino beam with significant contributions at
energies above 500\,MeV. The \cerenk\ detector is a spherical tank
of 40\,ft diameter filled with 776 tonnes of mineral oil viewed by
1275 8'' PMT's at a distance of 500\,m from the neutrino
production point. Again, the neutrino flux must be known
precisely by MC simulations and special checks. The expected
fluxes as well as a schematic view of the detector are shown in
Fig.~\ref{fig:booneflux}.
\begin{figure}[htb] \begin{center}
\includegraphics[width=14cm]{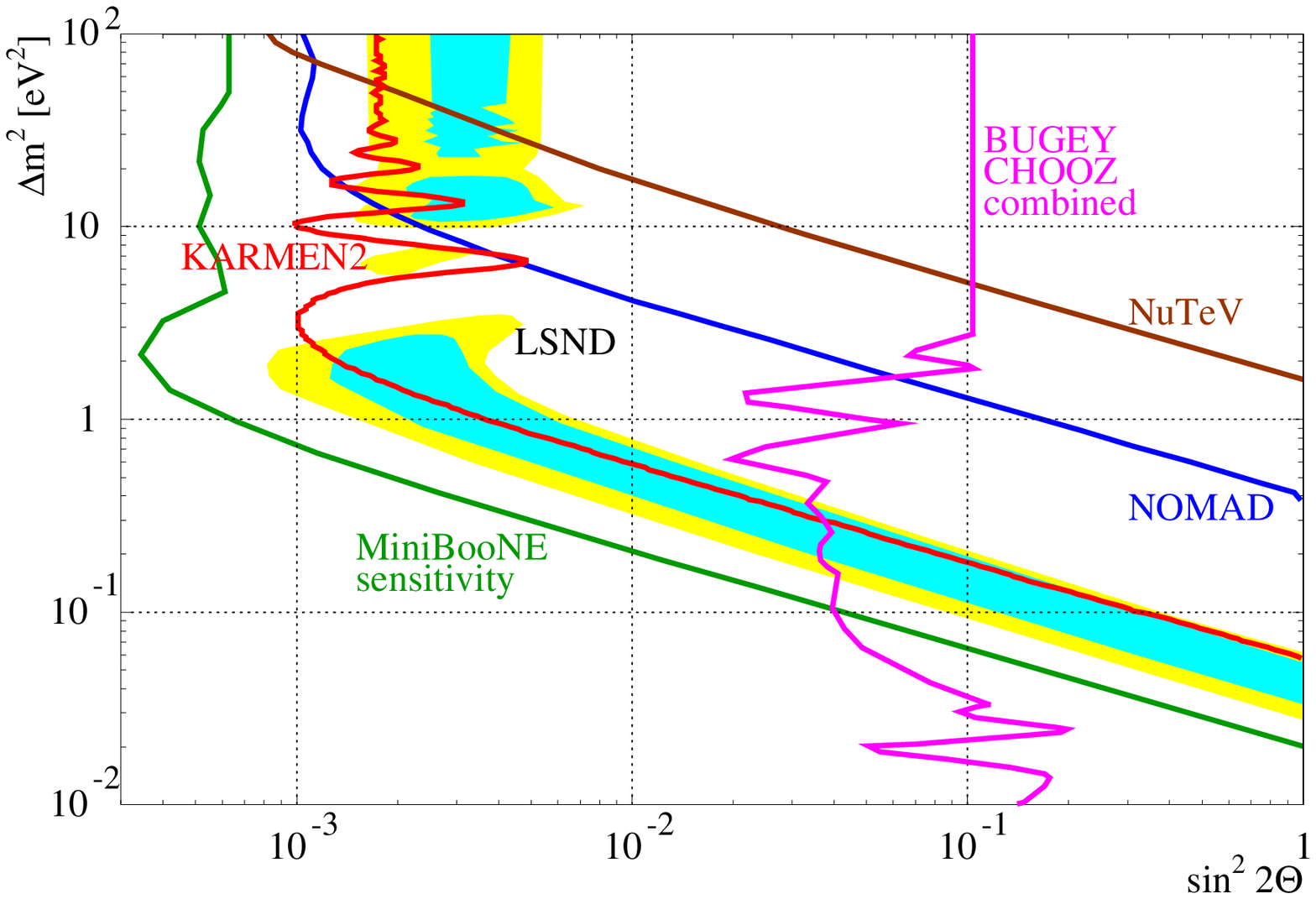}
 \caption{\it MiniBooNE expected \numunue\ sensitivity for the single horn
 design in comparison with the results of the above experimental searches,
 including the combined limit from the Bugey \protect\cite{Bugey}
 and CHOOZ \protect\cite{Chooz} reactor experiments searching for \nueb\
 disappearance.
 \label{fig:osciall} }
\end{center} \end{figure}
After 2 years of data taking corresponding to $10^{21}$ protons
on target, MiniBooNE expects to detect about $5\cdot 10^5$
\numu--induced CC events and 3500 electron-like events out of
which about 1000 events would arise from a \numunue\ oscillation
according to the LSND signal \cite{Rex}. The expected
experimental sensitivity is shown in Fig.~\ref{fig:osciall} as
exclusion curve in case there will be no oscillation signal. The
detector is already operational, the first neutrino beam is
scheduled to be delivered mid of 2002.

\section{Conclusions}\label{sec:Conclusions}

We have discussed the LSND evidence for \numubnueb\ oscillations
together with the negative searches performed by KARMEN, NOMAD
and NuTeV in the same flavor channel. The KARMEN result is the
most limiting one confirmed by NOMAD and NuTeV at higher values
of \Dm . Though there are no further hints than the LSND result
for neutrino oscillations with a mass difference \Dm\ in the
order or larger than 0.1\,eV$^2$, no existing short baseline
experiment can rule it out completely. The upcoming MiniBooNE
experiment should be sensitive enough to definitively confirm or
refute the LSND result interpreted as appearance of \nueb\ due to
oscillations \numubnueb .

\end{document}